# Monte Carlo Study of the Crystalline and Amorphous NaK Alloy

Doug Reitz and Estela Blaisten-Barojas

Computational Materials Science Center and
Department of Computational and Data Sciences
George Mason University, Fairfax, VA, USA
dreitz@gmu.edu, blaisten@gmu.edu

**Abstract**

Metropolis Monte Carlo simulations of the eutectic NaK alloy are performed using the Second Moment Approximation (SMA) model potential across a wide range of temperatures at constant pressure. The alloy structure and thermodynamics are analyzed along with the atomic level structures using a variety of structure identification methods. Both enthalpy and density are followed along an annealing process that reveals a clear melting point around 260 K. At lower temperatures, two thermodynamic branches are identified as crystalline and amorphous solids.

## 1 Introduction

Studying the properties of bulk alloys is of interest for a variety of potential applications. One area is the study of bulk metallic glasses (BMG) and other non-crystalline systems. While the identification of crystals is known, the characterization of glasses and amorphous structures is not as simple. New and improved simulation and analysis techniques to explore and analyze binary alloy configuration space have the potential to improve discovery in this area. Atomistic simulations have been shown to be useful in determining the structure of pure alkali-metal clusters modeled through a novel Second Moment Approximation (SMA) potential [1]. Experimental measurements of the mass density have been performed for the sodium-potassium, NaK, liquid phase from room to higher temperatures at the eutectic concentration [2]. Recently, the Na-K alloy at various relative concentrations was investigated [3] using the SMA potential of [1] for discovering the structure of nanoclusters. In this work it was noted that the 55-atom nanoalloy cluster gives formation to a core shell structure that correlates with the eutectic [3].

Eutectic NaK, 23% Na and 77% K in atomic weight, is an interesting system because the alloy is liquid at room temperature, thus, having applications as as a coolant in nuclear reactors among other systems [4]. The eutectic of a binary metal alloy is the composition thought to have the highest glass forming ability (GFA)[5], though other studies suggest near but off-eutectic compositions may have the highest GFA [6].

In this study, we conduct Metropolis Monte Carlo simulations with the SMA potential of the eutectic NaK alloy across a wide range of temperatures. The isobaric-isothermal ensemble NPT (constant number of atoms, pressure, and temperature) simulations are used to survey and explore the configuration space of the system. The resulting thermodynamic properties (density, enthalpy) are analyzed and followed along an annealing procedure that brings the system from high temperatures to low temperatures. Additional structure identification, pair correlation functions, and order parameters are used to analyze the sequence of characteristic equilibrium states corresponding to the liquid phase and the two well defined solid phases that develop at temperature below the melting temperature.

## 2 Model and Methods

The Second Moment Approximation (SMA) model potential is a many-body potential that approximates the local environment of every atom. The SMA is a classical representation of the tight binding approach. The model potential is a sum of terms that mimic the distribution of electronic states in a d-band, $U_{el}$, supplemented with a repulsive Born-Mayer potential to account for the repulsion between electrons at short range, $U_{rep}$. Thus, the SMA analytical expression is a sum of two terms $U_{coh} = U_{el} + U_{rep}$ with

$$U_{rep} = \sum_{i=1}^{N} \varepsilon_0 \sum_{j=1, i \neq j}^{N} e^{-p(\frac{r_{ij}}{r_0} - 1)} \tag{1}$$

$$U_{el} = -\sum_{i=1}^{N} \Big\{ \sum_{j=1, i \neq j}^{N} {\zeta_0}^2 e^{-2q(\frac{r_{ij}}{r_0} - 1)} \Big\}^{1/2} \tag{2}$$

where $r_{ij}$ are the interatomic distances and $N$ is the total number of atoms. The parameters $\zeta_0, \varepsilon_0, p, q, r_0$, for pure potassium and sodium were reported in [1] and shown in Table 1. For the heteroatomic interactions, it has been shown that a variety of combination rules did not result in substantial differences in the structure of the system [3]. We select the geometric mean of the homoatomic values of Na and K interactions for the four parameters between sodium and potassium atoms $\zeta_0, \varepsilon_0, p, q$ such that $parameter_{NaK} = \sqrt{parameter_{Na} * parameter_K}$. Meanwhile, the parameter $r_0$ for the K-Na pairs affects the density of the final system. Therefore, both Lorentz combination rules, arithmetic mean and geometric mean, were considered in several simulations of the system in the liquid state. The most appropriate combination rule for the heteroatomic parameter $r_0$ was found to be a weighted arithmetic average where the weights are the relative atomic ratios. Thus, $r_{0NaK} = \frac{N_{Na}}{N} r_{0Na} + \frac{N_K}{N} r_{0K}$, where $N_K, N_{Na}$ are the number of potassium and sodium atoms, respectively. This combination rule for $r_0$ provided an equilibrium density for the simulated NaK eutectic alloy in close agreement with the experimental value [2]. Table 1 lists the heteroatom parameters used in this work.

Metropolis Monte Carlo (MMC) is an algorithm that allows to simulate system properties in thermodynamic equilibrium by generating a series of system states consistent with the Boltzmann distribution

$$P_i = exp(-E_i/k_B T) / \sum_j exp(-E_j/k_B T), \tag{3}$$

| | $\zeta_0$(eV) | $\varepsilon_0$(eV) | $p$ | $q$ | $r_0$(Bohr) |
|---|---|---|---|---|---|
| Na | 0.29113 | 0.015955 | 10.13 | 1.30 | 6.99 |
| K | 0.26259 | 0.020545 | 10.58 | 1.34 | 8.253 |
| NaK | 0.27649 | 0.018105 | 10.35 | 1.32 | 7.844 |

Table 1: Parameters of the SMA model potential entering in Eqs. 1,2 for the pairs K-K, Na-Na and Na-K.

where $P_i$ are the probabilities of finding the system in state $i$ with energy $E_i$ at temperature $T$ and $k_B$ is the Boltzmann constant. The denominator in Eq. 3 is the partition function of the system that acts as a normalization factor. In the MMC algorithm each sampled state is a system's configuration given by the coordinates of all atoms composing the system. The generated sequence of samples are linked through a Markov chain that requires ratios of probabilities between two consecutive samples as transition links between chain elements. This eliminates calculation of the denominator in Eq. 3. The acceptance/rejection choice for transitioning from state $i$ to state $j$ is given by $min(1, P_j/P_i)$ and the principle of detailed balance (transition probabilities $i \rightarrow j = j \rightarrow i$) [7].

Computer simulations were run for a system of 2000 atoms with periodic boundary conditions to simulate a bulk material with a cutoff radius of 45 Bohr. At the eutectic concentration of the alloy the computational box has 648 sodium atoms and 1352 potassium atoms. The SMA model potential is used to compute the potential energy of the atomic configurations. The MMC is started from a configuration where the sodium atoms are randomly distributed in the sites of a perfect bcc lattice. The remaining sites are populated with potassium atoms. A new system configuration is generated by moving a single atom a random distance and direction. The magnitude of this atomic movement is dynamically adjusted throughout the simulation to maintain a 50 % rejection rate of attempted atomic moves.

The NPT version of the MMC was run where N, T and pressure are kept constant such that the probability of the system transitioning from state $i$ to state $j$ is given by

$$P_{i \rightarrow j} = exp\Big[ - [(E_j - E_i) + P(V_j - V_i)]/k_B T + N ln(V_j/V_i) \Big] \quad (4)$$

where $V_j, V_i$ are the volumes of the $j, i$ states, respectively. In the simulations the volume change of the computational box was attempted once 2000 atoms had attempted a move at constant volume.

Typically, in order to obtain the average values reported in the next section, two million passages through the full 2000 atoms were attempted after the system was sought to be in equilibrium.

# 3 Results and Analysis

Initially, the system was equilibrated at a high temperature of about 700 K. At these temperatures the system is a hot liquid. Next the system was slowly cooled down by taking the last configuration of the higher temperature to be the starting configuration at the lower temperature. The equilibrium density obtained at 350 K is 0.89 g/cm$^2$, which compares well with the experimental value of 0.85 g/cm$^2$ [2]. It is not surprising that the density from the simulation is slightly higher than the experimental value since the optimal lattice constant of the pure metals was less than the experimental [1].

During the cooling process and close to the melting temperature, the cooling rate was accelerated to avoid crystal formation and allow formation of an amorphous branch and a crystalline branch. In addition, once several points were obtained for the crystalline branch at lower temperatures of about 160 K, these configurations were slowly and gradually warmed up to produce crystalline configurations at higher temperatures. With this gradual warming, 239 K was the highest temperature at which the system remained crystalline for 2,000,000 MMC steps. However, averages between 241 K and 260 K have some configurations that correspond to the crystalline system. This is expected, since during a phase transition the system oscillates between two phases. Figure 1 shows results for the average density and enthalpy per atom at each attempted temperature. Averages are depicted as liquid, amorphous, and crystalline based on the density and enthalpy obtained at that temperature and the structural analysis performed for each temperature. The transition to the amorphous solid is very gradual, spanning the 260 K to 160 K temperature range of the supercooled liquid where solid and liquid regions coexist.

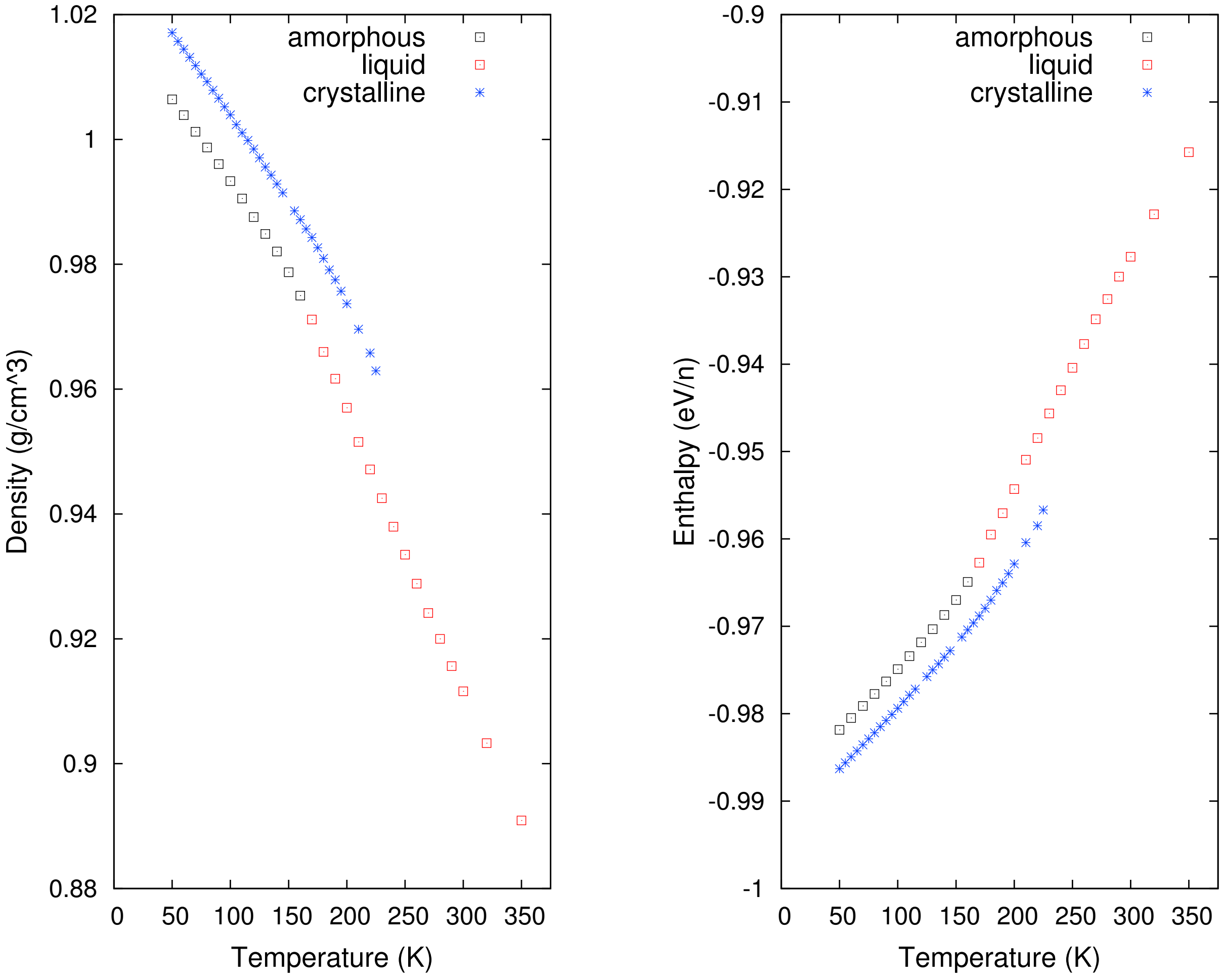

Figure 1: Density and enthalpy of the NAK eutectic alloy as a function of temperature. Red points depict the liquid region, black points correspond to the amorphous solid, and blue points pertain to the crystalline solid.

The two solid branches have different structural characteristics that are revealed by an analysis of the pair correlation function $g(r)$. To obtain pair correlation function data, configurations of the amorphous and crystalline systems at 120K where selected. Subsequently, a MMC-NVT run was started from each of them and the corresponding $g(r)$ was computed as an average over those new runs. The $g_{K-K}(r)$ are shown in Fig.2 left. A clear difference between the crystalline and amorphous systems is seen in the 2nd through 4rd peaks and in the general loss of structure with increasing r in the amorphous compared to the crystalline solids. The peaks in the crystalline solid match well with the bcc lattice. The definition of the crystal peaks is not full because the potassium atoms occupy only 68% of the sample volume. The remaindering sample volume is occupied by one cluster containing most of the sodium atoms. Sodium atoms in this cluster are also organized in an incipient bcc structure as shown by the $g_{Na-Na}(r)$ in Fig. 2 right. However, the Na cluster is extended, as desiring to form a lamellar structure, and thus presenting a large surface area with expected lattice reconstruction. By contrast, the amorphous system at this low temperature contains three Na clusters and a significantly larger number of isolated Na atoms. The structure within the Na clusters of the amorphous solid is also amorphous.

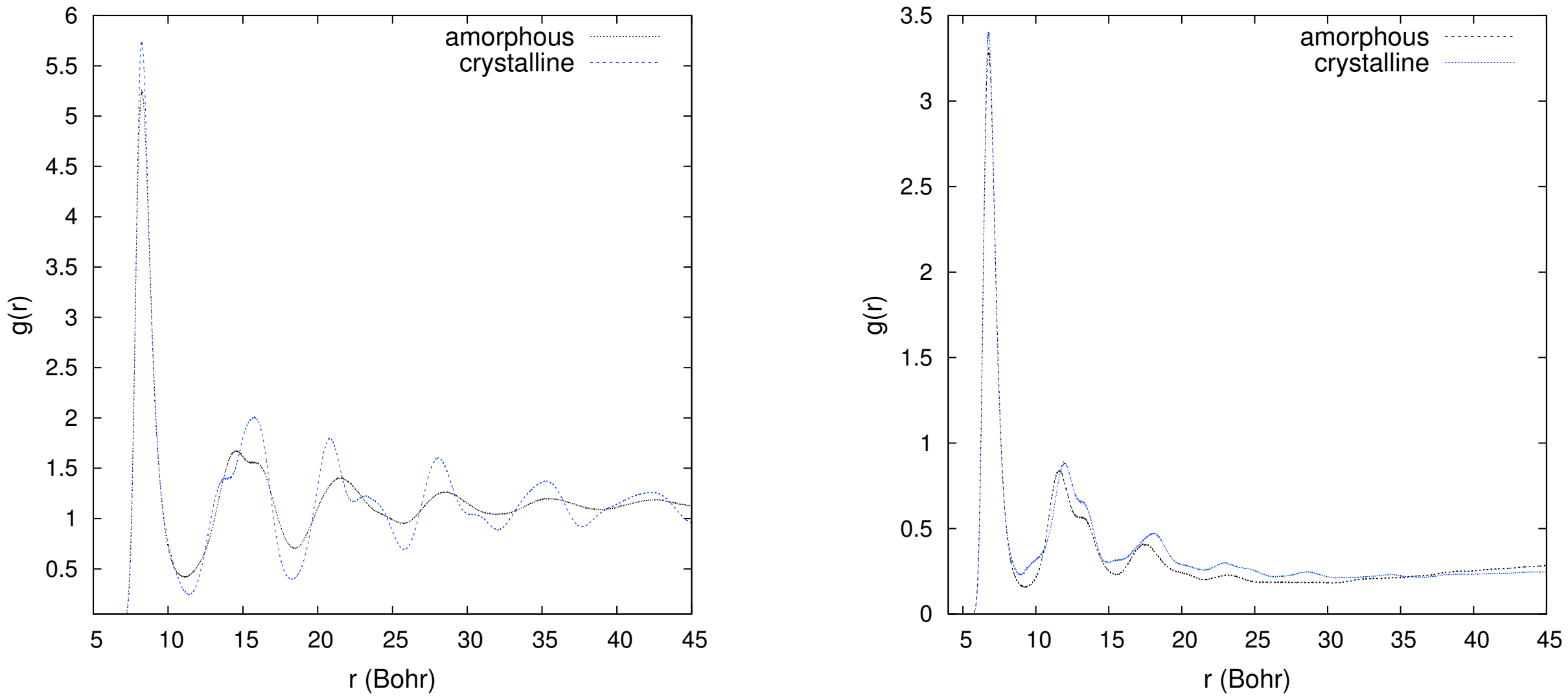

Figure 2: Pair correlation function for the K-K (left) and Na-Na (right) contributions

Figure 3 is a view of configurations at 120 K of the two solid systems pictorially showing differences between them. The amorphous branch cannot be identified as a glass because there has been formation of clusters of the minority atomic component instead of having a fairly homogeneous distribution of sodium atoms across the sample.

A useful structural order parameter is the ratio of the peak height between the first peak and subsequent peaks [7]. Shown in Fig. 4 are these ratios for the two solid samples as compared to a perfect bcc at the density of the crystalline sample. The first two ratios of $g_{K-K}(r)$ peaks in the crystalline sample display the same behavior than a perfect bcc while the two first points of the amorphous sample have a different behavior. As a result, we assert that the local environment up to the third coordination shell of the crystalline solid has bcc characteristics. The difference between the crystalline and amorphous samples are evident.

Another analysis of the atomic structure of the two solids is done with the Common Neighbor Analysis (CNA) algorithm [8, 9] that computes a fingerprint for pairs of atoms. Having two atom types requires the use of the Adaptive CNA (a-CNA) method [10]. This method determines the optimum cutoff radius of the coordination shell for each type of atom separately. Results from the a-CNA (as implemented in the Ovito package [11]) are given in Table 2. The crystalline solid manifests a high number of bcc atoms with about 20 % of other crystal symmetries due to boundary atoms between the K matrix and the Na cluster. On the other hand, the amorphous solid shows insignificant number of atoms that retain a local crystal environment. In summary, the structural analyses provide support that the solid samples obtained below the meting temperature can be differentiated not only by their thermodynamic behavior, but also by their structural signatures. The crystalline branch contains a matrix of potassium atoms that span the computational box and encase an extended cluster of sodium atoms. Meanwhile, the amorphous solid displays an amorphous matrix of potassium atoms encasing several clusters of sodium atoms that are themselves also amorphous. The characteristics of the amorphous solid discourage identification with a glassy solid.

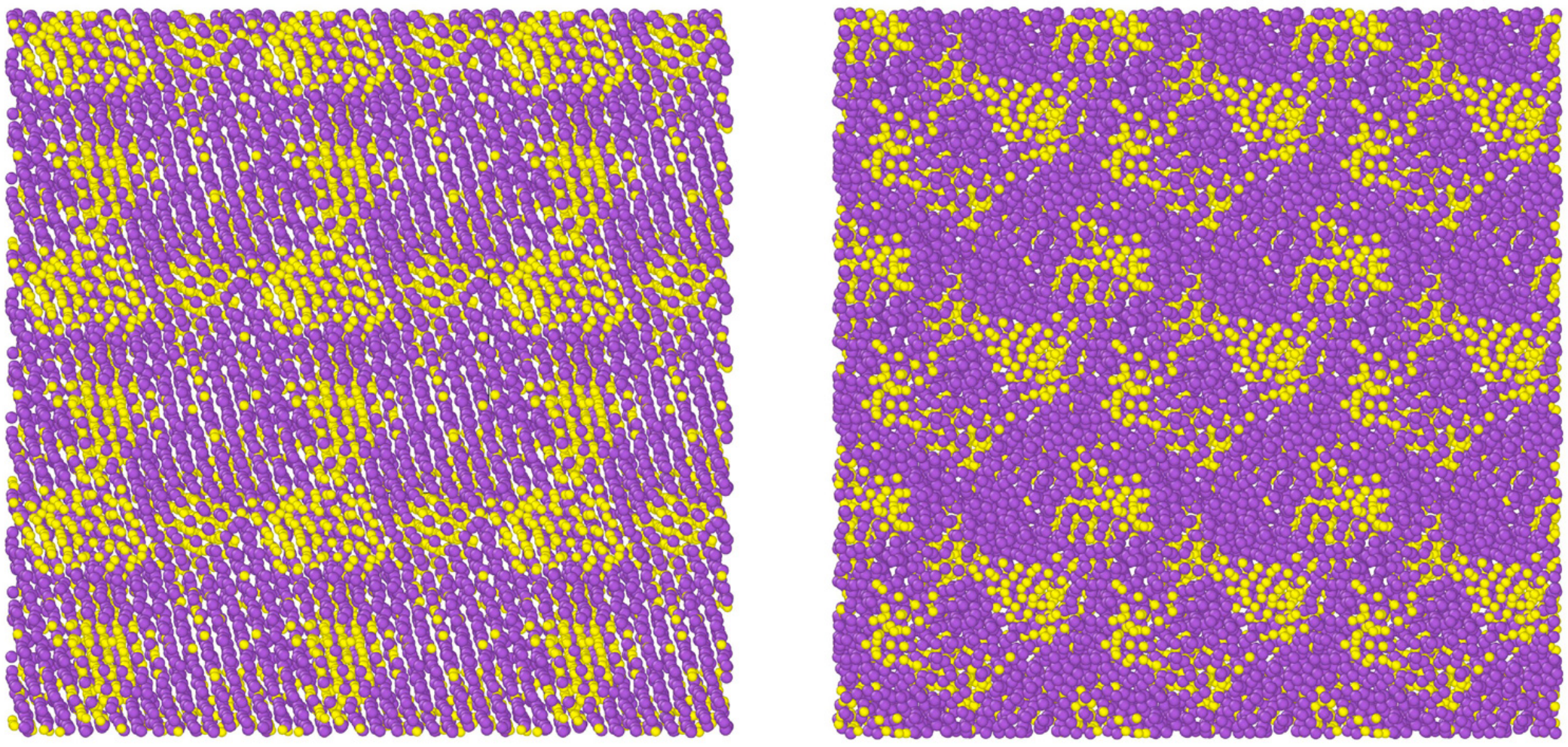

Figure 3: Snapshot of configurations of the crystalline (left) and amorphous (right) solids at 120 K.

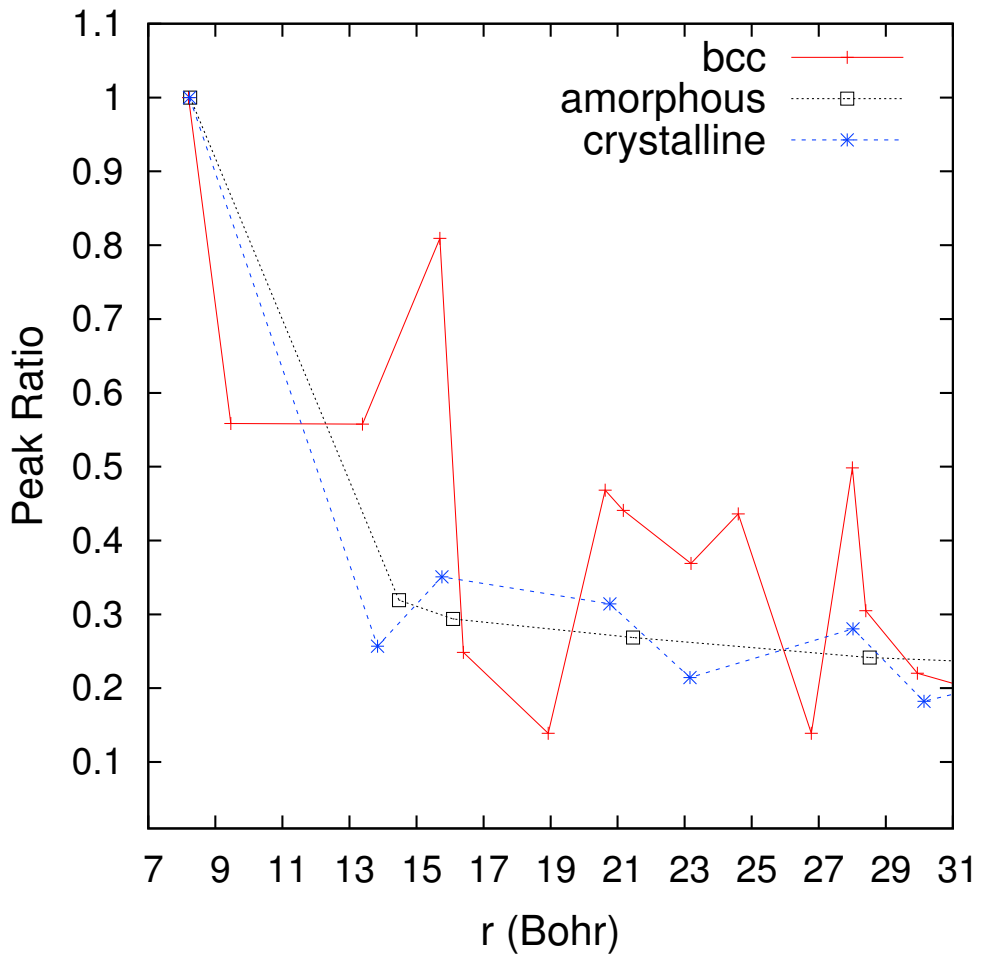

Figure 4: Order parameter obtained form the peak height ratios of the $g_{K-K}(r)$ compared to the perfect bcc values.

| Structure | Amorphous | Crystalline |
|---|---|---|
| Other | 1984 | 1320 |
| FCC | 2 | 72 |
| HCP | 6 | 60 |
| BCC | 6 | 548 |
| ICO | 2 | 0 |

Table 2: Adaptive common neighbor analysis of the two solid samples at 120 K

# 4 Conclusions

This study shows that MMC-NPT simulations with the SMA model potential of the NaK eutectic alloy predicts well an alloy melting region around 250-260 K and displays two solid branches below the melting temperature, a crystalline solid and an amorphous solid. The enthalpy and density as a function of temperature, as well as the structural properties of the two types of solids, ensure their profound differences. The crystalline solid is composed of a matrix of potassium atoms that span the computational box and has a bcc crystal structure with an embedded extended cluster of sodium atoms where the bcc structure is not evident. The amorphous solid is composed of an amorphous matrix of potassium atoms with several embedded sodium clusters that are also amorphous. The structural characteristics observed in the amorphous solid are not consistent with a glassy solid having homogeneously distributed minority atoms.